%Paper: hep-ph/9403365
%From: RAGAZZON%VSTST1@BNLDAG.AGS.BNL.GOV
%Date: Fri, 25 Mar 1994 08:41:31 -0500 (EST)
%Date (revised): Wed, 30 Mar 1994 07:49:27 -0500 (EST)

\pageno=0
\baselineskip=15pt plus 2pt
\magnification=\magstep1
\hsize=5.9truein
\vsize=8.7truein
\null
\hfill {\bf Preprint UTS-DFT-94-04}
\vskip 0.2 truecm
\centerline{ \bf SEMI-HARD PARTON RESCATTERINGS }
\centerline{ \bf IN NUCLEAR COLLISIONS AT VERY HIGH ENERGIES}
\vskip .25in
\centerline{R. Ragazzon and D. Treleani}
\vskip .05in
\centerline{\it Dipartimento di Fisica Teorica dell'Universit\`a
and INFN Sezione di Trieste}
\centerline{\it Trieste, I 34014 Italy}
\vfill
\centerline{ABSTRACT}
\vskip .25in
\midinsert
\narrower\narrower
\noindent
Representing the semi-hard partonic interactions
by the exchange of  Lipatov's perturbative Pomeron,
we express the semi-hard nuclear
cross section as a self shadowing cross section.
With the help of a generating functional technique, we obtain
average numbers of multiple semi-hard partonic collisions
without any need of using explicit expressions for the multi-parton
distributions.
The average number of semi-hard interactions of a
given projectile parton against a target nucleus is estimated
quantitatively
and it is shown to grow very rapidly above one with increasing
the c.m. energy.
\endinsert
\vfill\eject
\par	{\bf 1. Introduction}
\vskip.25in
One of the main problems in the physics of semi-hard
interactions is to take into account unitarity
corrections. The semi-hard regime, in fact,
although perturbative in the elementary interaction, gives rise
to cross sections
that may violate the unitarity limit.
Directly related features are the large size of the QCD-parton model
cross section to produce large $p_t$ partons, which is
obtained when the cut-off $p_t^{min}$, defining the
perturbative regime, is lowered down to the
mini-jet region[1], and the large value of
the mini-jet cross section, which has been observed by UA1[2].
The standard way to unitarize the semi-hard cross section is
by eikonalizing the elementary semi-hard
partonic interaction[3]: The semi-hard component of the hadronic
interaction is included in the eikonal phase as an additive term,
which represents the inclusive cross section to produce
large $p_t$ partons, integrated down to
the lower limit $p_t^{min}$ and depending explicitly on the
impact parameter of the hadronic collision.
The cross section, including
all inelastic events with at least one semi-hard partonic interaction,
may be easily derived and expressed as a series of multiple semi-hard
partonic interactions. The resulting distribution in multiple
semi-hard partonic interactions is a Poissonian,
with average number depending on the impact parameter and
representing different pairs of
partons interacting independently at different points of the transverse plane.
The underlying physical assumption is a Poisson distribution for the
input multi-parton distributions to the multi-parton interaction and,
in addition, the absence of any semi-hard
rescattering for each parton which has interacted already
once with semi-hard momentum transfer exchange[4].
\par
To include semi-hard interactions
in high energy nuclear collisions,
the eikonalized expression of the nucleon-nucleon interaction
can be used as a input
to construct the nuclear interaction[5]
according with Glauber and Gribov[6]. With respect to the input parton
distributions to the semi-hard component of the nuclear
interaction,
both possibilities of input nuclear parton distributions, with
and without shadowing corrections, have been considered[5].
Alternatively[7], the whole semi-hard nuclear interaction
has been expressed by a Poisson distribution of multiple semi-hard
parton collisions, as in the hadron-hadron case.
The only difference, with respect to the case of
semi-hard hadron-hadron interaction,
is in the function of the impact parameter,
which describes the overlap of the matter distribution in the
transverse plane, and in the nuclear parton distributions, which,
for nucleus-nucleus, include shadowing corrections[8]. In both of these
approaches a parton
is allowed to interact, with momentum transfer larger than $p_t^{min}$,
only once. Semi-hard parton rescatterings, namely multiple
collisions of a given parton, each with momentum transfer larger
than $p_t^{min}$, are
typically estimated to be a negligible effect up to energies
of order of $10TeV$ in the nucleon-nucleon c.m. system[8].
In this respect the very different conclusion, that semi-hard parton
rescatterings are, on the contrary, one of the major features in nuclear
collisions
already at energies of the order of $1TeV$ in the nucleon-nucleon
c.m. system, has been drawn[9].
This different conclusions is the origin of the rather peculiar
physical picture of high energy nuclear interactions
where most of the transverse energy
produced is the result of multiple semi-hard parton collisions[10].
\par	The purpose of the present note is to gain a better insight
into this last point; namely the amount of semi-hard
parton rescatterings in very high energy nuclear collisions,
which appears as a critical element in the physics of nuclear
interactions at very high energy.
We will in fact review the arguments of ref.[9, 10] making clear
which are the underlying assumptions
and, by means of an approach which is much more general with respect to the
approach
used there, we will produce estimates of the average number of semi-hard
rescatterings which are of much wider validity. In fact our result will not
depend
on any specific form for the input multi-parton distributions.
\par	The paper is organized in two parts:
In the first we comment on the general framework
to discus the semi-hard cross section and
we derive the quantities of interest for the actual discussion.
In the second part we present a few
numerical results and our conclusions.
\vskip .25in
\par    {\bf 2. General framework and average number of rescatterings}
\vskip .15in
\par
The inelastic channels are obtained from
the nuclear cross section, expressed according with
Glauber and Gribov, by
the use of the cutting rules by Abramovskii, Gribov and
Kancheli[11].
The rules connect different cut diagrams, which are
shown to be proportional to one another, in such a way that the evaluation of
the
contribution of the various inelastic channels to the inelastic cross section
simplifies
enormously. Actually the result, which is obtained
after summing all different cut diagrams,
is that the cross section acquires the simple probabilistic meaning
of sum of multiple inelastic interaction probabilities[12].
Remarkably the probabilistic picture is the quantum mechanical
result of the sum of a very complicated set of cut amplitudes.
The systematic  use of the cutting rules
allows also to recognize the validity of general properties
of the interaction, such as the possibility to single out
the self shadowing cross sections[13]. For these
quantities the unitarity corrections have the peculiar
propriety that all contributions from all processes, different
from the one which is considered, cancel. A stronger
statement is that also the functional form of the unitarity
corrections for a self shadowing cross section is given,
once the forward nuclear amplitude is expressed
in terms of "elementary" nucleon-nucleon amplitudes[14].
The question of interest for the present discussion is whether semi-hard
interactions are quantities of this kind.
Two are the requirements which are to be satisfied.
The first is that the soft corrections to the semi-hard process have to
cancel, when all contributions, real and virtual, are taken into account.
The proof of factorization at the
level of power corrections[15] shows that this is the case when inclusive
processes are considered. The second requirement, which allows the
probabilistic picture of the interaction, is that the cutting
rules, relating different cut diagrams, are to be valid
also for semi-hard interactions. While this proof is
missing in the general case, it has nevertheless been obtained for
one of the components of the semi-hard cross section which is
leading in the high energy limit[16]. One may therefore
argue that it is a good working
hypothesis to assume that, in high energy nuclear collisions,
semi-hard interactions self shadow, in such a way that
the semi-hard nuclear interaction can be expressed in a probabilistic way
as a function of "elementary" semi-hard
parton-parton interaction probabilities.
The physical picture of the semi-hard nuclear interaction discussed
in ref.[9] and in ref.[10] and the
estimates of the rates of semi-hard parton rescattering,
which we are going to present here, are done under this basic hypothesis.
\par
The self-shadowing hypothesis and the probabilistic picture
which follows are not sufficient to write
the semi-hard nuclear cross section.
In fact, differently with respect to
soft interactions, where the number of
"elementary" interacting objects, which in that case
are the
nucleons, is fixed, when considering semi-hard interactions
and partons as elementary interacting objects,
the number is, on the contrary, varying. Actually the configurations
with many interacting partons involve the multi-parton
distributions[17] that are non perturbative quantities,
independent of the single-parton distributions of
large $p_t$ physics. The complete description of the nuclear
semi-hard interaction needs
therefore, as a input, the infinite set of non perturbative
quantities, represented by the multi-parton distributions.
Nevertheless, for the more limited program to evaluate a
few average quantities, much less information is needed as a input.
In this case it is very convenient to approach the problem by means of a
generating functional formalism which allows to
avoid explicit expressions for the multiparton
distributions[18].
\par
To construct the generating functional for the multi-parton
distributions one may start with the exclusive multi-parton
distributions, namely, at a given scale provided by the cut off $p_t^{min}$,
the probabilities for each of the configurations with
a given number of partons. The $W^{(n)}(u_1\dots u_n)$
are then the exclusive $n$-body parton
distributions, where $u_i\equiv(b_i,x_i)$ represents the transverse partonic
coordinate $(b_i)$
and longitudinal fractional momentum $(x_i)$.
The generating functional ${\cal Z}[J]$ is defined as:

$${\cal Z}[J]=\sum_n{1\over n!}\int J(u_1)\dots J(u_n)W^{(n)}(u_1\dots u_n)
du_1\dots du_n,\eqno(1)$$

\noindent
with the normalization condition

$${\cal Z}[1]=1.\eqno(2)$$

\noindent
While the exclusive distributions are obtained from
the expansion of the generating functional around $J=0$,
the inclusive distributions are obtained by expanding
${\cal Z}[J]$ around $J=1$. Explicitly the one-body  and
two-body inclusive distributions $D^{(1)}$ and $D^{(2)}$ are given by:

$$\eqalign{D^{(1)}(u)=&W^{(1)}(u)+\int W^{(2)}(u,u')du'+{1\over 2}
		    \int W^{(3)}(u,u',u'')du'du''+\dots\cr
		 =&{\delta{\cal Z}\over \delta J(u)}\biggm|_{J=1}
		 ,\cr
     D^{(2)}(u_1,u_2)=&W^{(2)}(u_1,u_2)+\int W^{(3)}(u_1,u_2,u')du'\cr
		    &+{1\over 2}
		    \int W^{(4)}(u_1,u_2,u',u'')du'du''\dots\cr
		 =&{\delta^2{\cal Z}\over \delta J(u_1)\delta J(u_2)}
		  \biggm|_{J=1}\equiv D^{(1)}(u_1)D^{(1)}(u_2)
      +{1\over 2}C^{(2)}(u_1,u_2)
}\eqno(3)$$

\noindent
where the two body correlation
$C^{(2)}(u_1,u_2)$, which measures the deviation from the
Poisson distribution[18], has been introduced.
\par    The semi-hard nucleus-nucleus
cross section $\sigma_H$ is constructed by multiplying the exclusive parton
distributions
of the two nuclei, $A$ and $B$, by the probability of interaction, which,
after the assumption that semi-hard interactions
self-shadow, can be constructed from the elementary
parton-parton interaction probability $\hat{\sigma}(u_i,u'_j)$, which
represents
the probability for the parton $i$
of the $A$-nucleus to have an hard interaction with the parton $j$ of the
$B$-nucleus:

$$\sigma_H=\int d\beta\sigma_H(\beta)$$

$$\eqalign{\sigma_H(\beta)=\int&\sum_n{1\over n!}
  {\delta\over \delta J(u_1)}\dots
  {\delta\over \delta J(u_n)}{\cal Z}_A[J]\cr
  \times&\sum_m{1\over m!}
  {\delta\over \delta J'(u_1'-\beta)}\dots
  {\delta\over \delta J'(u_m'-\beta)}{\cal Z}_B[J']\cr
\times&\Bigl\{1-\prod_{i=1}^n\prod_{j=1}^m\bigl[1-\hat{\sigma}(u_i,u'_j)\bigr]
   \Bigr\}\prod dudu'\Bigm|_{J=J'=0}}
\eqno(4)$$

\noindent
Here $\beta$ is the impact parameter between the two interacting nuclei
and the semi-hard cross section is constructed by summing over all possible
partonic configurations of the two interacting nuclei (the sums over
$n$ and $m$) and, for each configuration with $n$ partons from $A$ and
$m$ partons from $B$, summing over all possible multiple partonic
interactions. This last sum is constructed by asking for the
probability of no interaction between the two configurations
(actually $\prod_{i=1}^n\prod_{j=1}^m[1-\hat{\sigma}_{i,j}]$ ). The
difference from one of the probability of no interaction
gives the sum over all
semi-hard interactions.
To obtain the average number of semi-hard partonic
collisions one expands the interaction probability
as a sum of multiple interactions.
Since the interaction
probability is multiplied by a symmetric expression,
one can make the replacement :

$$\eqalign{1-\prod_{i=1}^n\prod_{j=1}^m[1-\hat{\sigma}_{i,j}]
  \to{\cal S}&[1-\prod_{N=1}^Q\hat{\sigma}_{N}]\cr
  ={\cal S}&\sum_{N=1}^Q
  {Q\choose N}\hat{\sigma}_1\dots\hat{\sigma}_{N}
  (1-\hat{\sigma}_{N+1})\dots(1-\hat{\sigma}_Q)
}\eqno(5)$$

\noindent
where the index $N$ counts the possible
interactions, in such a way that $Q=nm$, ${\cal S}$
is a symmetrizing operator[9], and, in the second line
in Eq.(5),  the interaction probability has been
expressed as a sum of multiple interactions. The average number of
partonic collisions at a given value of $Q$, $\langle N\rangle_Q$, is given by:

$$\eqalign{\langle N\rangle_Q&={\cal S}\sum_{N=1}^QN
  {Q\choose N}\hat{\sigma}_1\dots\hat{\sigma}_{N}
  (1-\hat{\sigma}_{N+1})\dots(1-\hat{\sigma}_Q)\cr
  &={\cal S}Q\hat{\sigma}_1
}\eqno(6)$$

\noindent
When in the expression for $\sigma_H(\beta)$, as given in Eq.(4),
the interaction probability is replaced with $\langle N\rangle_Q$,
the overall average number of parton interactions $\langle N(\beta)\rangle$ is
obtained.
Since ${\cal S}Q\hat{\sigma}_1=mn\hat{\sigma}_{1,1}$,
as a result of the the sums on $m$ and $n$
both arguments of
${\cal Z}_A$ and of ${\cal Z}_B$ are
shifted by one unit. One can then write:

$$\langle N(\beta)\rangle=\int
{\delta\over \delta J(u_1)} {\cal Z}_A[J+1]\Bigm|_{J=0}
{\delta\over \delta J'(u_1'-\beta)}
{\cal Z}_B[J'+1]\Bigm|_{J'=0} \hat{\sigma}(u_1,u_1')
du_1du_1'\eqno(7)$$

\noindent
The overall average number of partonic collisions is therefore
expressed, on quite general grounds, by the single scattering term, where
the parton structure of each interacting nucleus enters
only with the inclusive one-body parton distribution:

$$\langle N(\beta)\rangle=\int D^{(1)}_A(u_1)
  \hat{\sigma}(u_1,u_1')D^{(1)}_B(u_1'-\beta)
  du_1du_1'\equiv D^{(1)}_A(u_1)\otimes\hat{\sigma}_{1,1}
  \otimes D^{(1)}_B(u_1'-\beta)\eqno(8)$$

\par
An analogously general result can be obtained for the
average number of collisions of each parton. The sum over $m$
in Eq.(4) can be performed explicitly in such a way that
$\sigma_H(\beta)$ is expressed as:

$$\eqalign{\sigma_H(\beta)=\int&\sum_n{1\over n!}
  {\delta\over \delta J(u_1)}\dots
  {\delta\over \delta J(u_n)}{\cal Z}_A[J]\cr
\times&\Bigl\{1-\prod_{i=1}^n{\cal Z}_B\bigl[1-\hat{\sigma}(u_i,\cdot)\bigr]
   \Bigr\}\prod_{i=1}^n du_i\Bigm|_{J=0}}
\eqno(9)$$

\noindent
In Eq.(9) every configuration with a given number of partons from the $A$
nucleus
is required to interact at least once with nucleus $B$.
The probability for one
of the partons form $A$, which we label with $i$, not to interact with the
whole
nucleus $B$ is ${\cal Z}_B\bigl[1-\hat{\sigma}(u_i,\cdot)\bigr]$. One minus
this probability gives the probability for that parton
to interact at least once. Let us expand this interaction
probability as a sum on successive interactions:

$$\eqalign{
  1-{\cal Z}_B\bigl[1-\hat{\sigma}(u_i,\cdot)\bigr]&=
 \int\sum_m{1\over m!}
  {\delta\over \delta J'(u_1'-\beta)}\dots
  {\delta\over \delta J'(u_m'-\beta)}
  {\cal Z}_B[J']\Bigm|_{J'=0}\cr
\times&\sum_{k=1}^m
  {m\choose k}\hat{\sigma}_{i,1}\dots\hat{\sigma}_{i,k}
  (1-\hat{\sigma}_{i,k+1})\dots(1-\hat{\sigma}_{i,m})
\prod du'
}\eqno(10)$$

\noindent
the average number of interactions of the parton under
consideration is obtained by multiplying by $k$ each term
in the sum on $k$. Both sums on $k$ and on $m$ can be done
explicitly in a way similar to the case already considered. The
resulting average number is expressed as:

$$\langle k(u_i)\rangle=\int
  D^{(1)}_B(u_1'-\beta)\hat{\sigma}(u_i,u_1')
  du_1'
  =\hat{\sigma}_{i,1}\otimes D^{(1)}_B(u_1'-\beta)
\eqno(11)$$

\noindent
where the structure of the target nucleus enters only with $D^{(1)}_B$.
\par    One can notice that, while the average
number of interactions of the parton with the target does
not feel more than the inclusive one-parton distribution
of the target, if one looks instead to the average number of
rescatterings of the same parton, one is sensible to
the multi-parton correlations.
Let us in fact evaluate the average number of
rescatterings $\langle r\rangle$, namely the average
number of interactions
when the number of interactions is at least two.
For a fixed configuration with $m$
target partons one needs to compute:

$$\eqalign{
  \langle r(u_i)\rangle_m=&\sum_{k=2}^mk
  {m\choose k}\hat{\sigma}_{i,1}\dots\hat{\sigma}_{i,k}
  (1-\hat{\sigma}_{i,k+1})\dots(1-\hat{\sigma}_{i,m})\cr
  =&m\hat{\sigma}_{i,1}-m\hat{\sigma}_{i,1}
  (1-\hat{\sigma}_{i,2})\dots(1-\hat{\sigma}_{i,m})
}\eqno(12)$$

\noindent
The average number of rescatterings $\langle r(u_i)\rangle$, for an incoming
parton
with a given kinematical configuration represented by $u_i$, is
obtained by multiplying the expression in Eq.(12) by the sum on $m$, giving
all the different multi-parton configurations, and by performing the sum.
The actual result is:

$$\langle r(u_i)\rangle=\hat{\sigma}_{i,1}
  {\delta\over\delta J'}\bigl[{\cal Z}_B[J'+1]
  -{\cal Z}_B[J'+1-\hat{\sigma}_{i,\cdot}]
  \bigr]\Bigm|_{J'=0}\eqno(13)$$

\noindent
By expanding ${\cal Z}_B[J'+1-\hat{\sigma}]$ for $\hat{\sigma}$
small and keeping only the first term
different from zero one obtains:

$$\langle r(u_i)\rangle=\hat{\sigma}_{i,1}\otimes
  \hat{\sigma}_{i,2}\otimes
  \Bigl[D^{(1)}_B(u_1'-\beta)D^{(1)}_B(u'_2-\beta)
    +{1\over 2}C^{(2)}_B(u_1'-\beta,u_2'-\beta)\Bigr]\eqno(14)$$

\noindent
in such a way that combining $\langle r(u_i)\rangle$
with $\langle k(u_i)\rangle$
the correlation $C^{(2)}_B$ is obtained.
\par    A similar observation is valid for $\langle N(\beta)\rangle$:
By starting the sum in Eq.(6) from $N=2$, one identifies
the average number $\langle\nu(\beta)\rangle$, which
counts those semi-hard collisions with
at least two semi-hard parton interactions.
At the lowest order in $ \hat{\sigma}$,
$\langle\nu(\beta)\rangle$ is expressed as:

$$\langle\nu(\beta)\rangle=D^{(1)}_A\otimes\hat{\sigma}_{1,1}\otimes
  \hat{\sigma}_{1,2}\otimes D^{(2)}_B+
 D^{(2)}_A\otimes\hat{\sigma}_{1,1}\otimes
  \hat{\sigma}_{2,1}\otimes D^{(1)}_B+
D^{(2)}_A\otimes\hat{\sigma}_{1,1}\otimes
  \hat{\sigma}_{2,2}\otimes D^{(2)}_B\eqno(15)$$

\noindent
In Eq.(15) the dominant term is obtained by neglecting the
correlations in the expression $D^{(2)}(u_1,u_2)=D^{(1)}(u_1)D^{(1)}(u_2)
      +1/2C^{(2)}(u_1,u_2)$
and keeping only the term with the larger number
of $D^{(1)}$'s:

$$\langle\nu(\beta)\rangle\approx
D^{(1)}_A\otimes\hat{\sigma}_{1,1}\otimes D^{(1)}_B
D^{(1)}_A\otimes\hat{\sigma}_{2,2}\otimes D^{(1)}_B\eqno(16)$$

\noindent
which represents the independent
semi-hard scattering of two uncorrelated parton pairs
localized at two different points in the transverse plane, as it can be
seen by looking at the dependence on the transverse coordinates in Eq.(16).
Actually this is the expression that has been used first to make
predictions on the
rate for double parton collisions, in high energy hadronic
interactions[19], and later to analyze the actual experimental signal[20].
In Eq.(15) the most important term containing correlation is:

$$D^{(1)}_A\otimes
    D^{(1)}_A\otimes\hat{\sigma}_{1,1}\otimes
  \hat{\sigma}_{2,2}\otimes{1\over 2}C^{(2)}_B+A
  \leftrightarrow B$$

\par\noindent
One may notice
that $\langle r(u_i)\rangle$ and $\langle\nu(\beta)\rangle$ give a different
information about the correlation. In the first case the transverse
coordinate of the two correlated partons differ only by an amount
whose scale is given by the range of $\hat{\sigma}$, as it is seen by looking
at the
convolutions in Eq.(14). In the second case the
correlation is averaged over the whole transverse plane.
\vfill
\eject
\par    {\bf 3. Numerical estimates}
\vskip .15in
\par
The analysis of the previous paragraph shows that, although multiple
parton collisions involve a whole infinite set of unknown non-perturbative
inputs, namely the multi-parton distributions, to construct average quantities,
only a limited amount of input information is needed. Actually
the average number of partonic interactions in
an inelastic event with given impact parameter $\beta$,
$\langle N(\beta)\rangle$, and the average number of
semi-hard interactions of a projectile parton, with given
transverse coordinate and fractional momentum
$u_i$, against a target nucleus,
$\langle k(u_i)\rangle$, are both constructed
from the one-body inclusive parton distribution $D^{(1)}$.
\par
In order to perform any estimate we need, as a input, $D^{(1)}$.
When the "elementary" interaction
probability is small, semi-hard rescatterings can be neglected.
In this case the integral on $\beta$ of $\langle N(\beta)\rangle$
is equal to the semi-hard cross section multiplied by the multiplicity
of partons which have suffered a semi-hard interaction[4], and, therefore,
it is equal to the inclusive cross section. As a consequence, in this case one
is
allowed to identify $D^{(1)}$ with the nuclear structure function
which is used to describe large $p_t$ physics. Notice that this conclusion
holds
at any order in the parton correlations of the multi-parton
inclusive distributions. It is therefore valid for the particular case of the
Poissonian
distribution in multiplicity of the
semi-hard parton collisions, which
corresponds to the limit of neglecting all correlations[18].
The identification of $D^{(1)}$ is much less clear when semi-hard parton
rescatterings are taken into account.
If looking to deep inelastic scattering on a nucleus, in the kinematical regime
of shadowing corrections, the fluctuation of the virtual photon
in a $q{\bar q}$ pair lasts long enough to cross the whole nucleus[21].
In this case, the interaction of the photon is
hadronic, or, more precisely, it can be described as the sum of
$q{\bar q}$ states, with frozen transverse distance
interacting
according to the Glauber expression for hadron-nucleus
amplitude in the forward direction[21].
On the other hand, in the Glauber expression, shadowing is the result of
all multiple rescatterings of the incoming state, both
soft and semi-hard. When the amount of semi-hard rescatterings
is sizeable, the problem which rises is that the quantity entering
in the total cross section is the probability
to interact at least once, while the quantity of interest for the present
purpose is rather the average number of interactions. Nuclear structure
functions, with shadowing corrections, are therefore
an underestimate for $D^{(1)}$. On the other hand nuclear structure
functions without shadowing corrections are an overestimate,
since also soft interactions alone produce shadowing at low $x$.
\par
To have a quantitative feeling of the importance of semi-hard rescatterings,
we compute $\langle k(u_i)\rangle$, as given in Eq.(11), using
as a input for $D^{(1)}$ both nuclear
structure functions with and without shadowing corrections.
The two choices represent a lower and an upper bound
for $\langle k(u_i)\rangle$. To perform the actual calculation we use the set E
and set B structure
functions by HMRS[22] and,
as scale factor, we consider two possible choices,
$p_t^{min}/2$ and $p_t^{min}$. Shadowing corrections
have been introduced by parametrizing the ratio of the nuclear-nucleon
structure functions as in ref.8.
\par
As a further input, we need to specify the elementary
interaction. The regime of interest is the one of mini-jet production.
In the typical configuration two minijets, namely
jets with
transverse momentum of a few Gev, have a large separation in rapidity.
As one expects from asymptotic estimates[23],
the rapidity interval is filled with radiated gluons, in such a way that
higher order corrections in $\alpha_S$ play an important role.
Asymptotically one obtains that, in the
inclusive cross section, the first correction to the
lowest order result is of order $(\alpha_Sy)^2$[24], with
$y$ the rapidity interval between the two interacting partons.
To keep into account the main features of higher order corrections,
we parametrize the "elementary" partonic
interaction by the exchange of the Lipatov's perturbative
Pomeron[23]. More explicitly we use the expression of the inclusive cross
section
to produce mini-jets derived
by Mueller and Navelet[24].
\par	The
inclusive cross section to produce mini-jets
with transverse momentum larger than $5GeV$, in
$p{\bar p}$ collisions with c.m. energy
range $200 GeV\le\sqrt s\le900GeV$, is compared
with the cross section measured by UA1[2] in fig.1.
Using as a input HMRS(B) structure functions and $p_t^{min}/2$ as scale factor
(continuous curve in the figure) the gross figures of the data are reproduced.
The other choices, which we have considered,
underestimate the cross section by roughly a factor two.
The average number
of interactions, that a given parton undergoes with the
target nucleus, $\langle k(b, x)\rangle$, as expressed by Eq.(11), is shown in
fig. 2
as a function
of the fractional momentum of the projectile parton $x$ and at
zero impact parameter $b$. The two continuous curves refer
to the two different choices of the cut off, $p_t^{min}=5GeV$
and $p_t^{min}=6GeV$, together with HMRS(B) structure functions
and $p_t^{min}/2$ as scale factor, which, according with the UA1 data, is the
favorite choice for the input parameters. The dependence on the scale factor
is shown by the dashed line in the same figure, where the input is
$p_t^{min}=5GeV$, HMRS(B)
structure functions and $p_t^{min}$ as scale factor.
With the same input, the effect of using structure functions, with shadowing
corrections
included, is shown by the dotted line. The dependence
on the c.m. energy $\sqrt s$ is shown in fig.3, where
the value of the cut-off has been fixed to $p_t^{min}=5GeV$.
Solid, dashed, dotted lines refer to the three values for the fractional
momentum of the projectile parton $x=1$, $x=0.5$ and $x=0.25$ respectively.
In fig.4 we show the dependence on the impact
parameter at fixed $\sqrt s$ and $p_t^{min}$. In all cases
the atomic mass number is $A=208$ and the nucleus is
represented as an uniform sphere of radius $R$ with sharp boundaries.
\par    Our conclusion is that the average number of semi-hard collisions, that
an incoming parton undergoes with a target nucleus, is sizeable larger
than one, also keeping the lower cut-off $p_t^{min}$ as high as
$5 GeV$. We mainly limit our
analysis to the case of interactions between nuclei with
sub-energies of the order of $1TeV$ in the nucleon-nucleon
c.m. system. The input to the nuclear case is therefore constrained
by the comparison with
the available experimental result for the inclusive cross section
to produce mini-jets, as measured by UA1[2] in $p{\bar p}$ collisions at
similar energies.
A reason for the big effect, in the number of semi-hard parton
rescatterings, is in the large size of the mini-jet cross section which has
been
observed by UA1: almost $20mb$ at $900GeV$ c.m. energy in
$p{\bar p}$ collisions, with the cut-off $p_t^{min}=5GeV$.
Nevertheless an important role is played by the actual representation of the
"elementary"
partonic interaction. When the "elementary" partonic interaction is represented
by the
exchange of Lipatov's perturbative Pomeron, the large value
of the cross section observed by UA1 is the result of a relatively large
"elementary"
cross section rather than of a large flux of incoming partons. In a nucleus
the probability of semi-hard parton rescattering is therefore enhanced.
\vskip.25in
\par	{\bf Acknowledgements}
\vskip.15in
\par\noindent
Helpful discussions with G. Calucci are gratefully acknowledged. This work
has been partially supported by the Italian Ministry of University and of
Scientific
and Technological Research with the Fondi per la Ricerca
Scientifica-Universit\`a
di Trieste.
\vfill
\eject

\par    {\bf References}
\vskip .15in

\item{1.} G. Pancheri and Y. Srivastava {\it Phys. Lett.} {\bf B182}
(1986)199; S. Lomatch, F.I. Olness and J.C. Collins {\it Nucl. Phys.}
{\bf B317}(1989)617.
\item{2.} C. Albajar et al. {\it Nucl. Phys.} {\bf B309} (1988) 405.
\item{3.} L. Durand and H. Pi {\it Phys. Rev. Lett.} {\bf 58} (1987) 303;
A. Capella, J. Tran Thanh Van and J. Kwiecinski, {\it ibid.} (1987) 2015.
\item{4.} Ll. Ametller and D. Treleani, {\it Int. J. Mod. Phys.} {\bf A 3}
(1988) 521.
\item{5.} F. W. Bopp, R. Engel, I. Kawrakow, D. Petermann and J. Ranft,
preprint INFN/AE-93/24, Nov. '93; I. Kawrakow, H.-J. M\"ohring
and J. Ranft, {\it Phys. Rev.} {\bf D47} (1993) 3849 and {\it Z. Phys.}
{\bf C56} (1992) 115.
\item{6.} R.J. Glauber, lectures in {\it Theoretical Physics, Vol. I}
(ed. W.E. Brittin {\it et. al.}), Interscience, New York (1959), p. 315;
V.N. Gribov, {\it Zh. Eksp. Theor. Fiz.} {\bf 56} (1969) 892; {\bf 57}
(1969) 1306 [ {\it Sov. Phys. JETP} {\bf 29} (1969) 483; {\bf 30} (1970)
709 ].
\item{7.} K. Kajantie, P.V. Landshoff and J. Lindfors, {\it Phys. Rev. Lett.}
{\bf 59}(1987) 2527;  K.J.  Eskola, K. Kajantie and J. Lindford,
{\it Nucl.Phys.} {\bf B323} (1989) 37; K.J. Eskola, {\it Z.Phys.} {\bf C51}
(1991) 633;
X. N. Wang and M. Gyulassy, {\it Phys. Rev.} {\bf D45} (1992) 844;
K.J. Eskola and Xin-Nian Wang, {\it Phys. Rev.} {\bf D49} (1994) 1284.
\item{8.} X. N. Wang and M. Gyulassy, {\it Phys. Rev.} {\bf D44}
(1991) 3501
\item{9.} G. Calucci and D. Treleani, {\it Phys. Rev.} {\bf D41} (1990)
3367.
\item{10.} G. Calucci and D. Treleani, {\it Phys. Rev.} {\bf D44} (1991) 2746;
in {\it Perturbative QCD and Hadronic Interactions}, Proceedings of the XXVIIth
Rencontre de Moriond, Les Arcs, Savoie, France, 1992, edited by J. Tran Thanh
Van
(Editions Frontiers, 1992); in {\it Multiparticle Dynamics 1992},
Proceedings of the XXII International Symposium on Multiparticle
Dynamics, Santiago de Compostela, Spain, 1992, edited by C. Pajares
(World Scientific, Singapore, 1993).
\item{11.} V. Abramovskii, V.N. Gribov and O.V. Kancheli, {\it Yad. Fiz.}
{\bf 18} (1973) 595 [{\it Sov. J. Nucl. Phys.} {\bf 18} (1974) 308 ].
\item{12.} J.H. Weis, {\it Acta Physica Polonica} {\bf B7} (1976) 851;
A. Capella and A. Krzywicki, {\it Phys. Lett.} {\bf 67B} (1977) 84;
{\it Phys. Rev.} {\bf D18} (1978) 3357.
\item{13.} R. Blankenbecler, A. Capella, C. Pajares, J. Tran Thanh Van and
A.V. Ramallo, {\it Phys. Lett.} {\bf 107B} (1981) 106; C. Pajares and A.V.
Ramallo,
{\it Phys. Rev.} {\bf D31} (1985) 2800.
\item{14.} K. Boreskov and A. Kaidalov, {\it Acta Physica Polonica}
{\bf B20} (1989) 397.
\item{15.} J. Qiu and G. Sterman, {\it Nucl. Phys.} {\bf B353} (1991) 137.
\item{16.} G. Calucci and D. Treleani, {\it Phys. Rev.} {\bf D49} (1994) 138.
\item{17.} H.D. Politzer, {\it Nucl. Phys.} {\bf B172} (1980) 349;
R.K. Ellis, R. Petronzio and W. Furmanski, {\it ibid.} {\bf B207} (1981) 1.
\item{18.} G. Calucci and D. Treleani, {\it Nucl. Phys.} {\bf B} (Proc. Suppl.)
18C (1990) 187 and
{\it Int. J. Mod. Phys.} {\bf A6} (1991) 4375.
\item{19.} N. Paver and D. Treleani, Nuovo Cimento {\bf A70}(1982)215; Zeit.
Phys.{\bf C28} (1985)187; B. Humpert, {\it Phys. Lett.} {\bf 131B} (1983) 461;
B. Humpert and R. Odorico, {\it ibid} {\bf 154B} (1985) 211; T. Sjostrand and
M. Van Zijl, {\it Phys. Rev.} {\bf D36} (1987) 2019.
\item{20.} F. Abe et al., {\it Phys. Rev.} {\bf D47} (1993) 4857.
\item{21.} N. N. Nikolaev and B. G. Zakharov, {\it Z. Phys.} {\bf C49} (1991)
607.
\item{22.} P.N. Harriman, A.D. Martin, W.J. Stirling and R.G. Roberts,
{\it Phys. Rev.} {\bf D42} (1990) 798.
\item{23.} E.A. Kuraev, L.N. Lipatov and V.S. Fadin, {\it Sov.  Phys. JETP}
{\bf 45} (1977) 199; Ya. Ya. Balitsky and L.N. Lipatov, {\it Sov. J. Nucl.
Phys. }
{\bf 28} (1978) 822.
\item{24.} A.H. Mueller and H. Navelet, {\it Nucl. Phys.} {\bf B282} (1987)
727.

\vfill
\eject
\par    {\bf Figure captions}
\vskip .15in

\item{Fig. 1} Inclusive cross section
to produce mini-jets
with transverse momentum larger than $5GeV$, in
$p{\bar p}$ collisions as a function of the c.m. energy.
Continuous line:
HMRS(B) structure functions, scale factor $p_t^{min}/2$. Dashed line:
HMRS(B), scale factor $p_t^{min}$. Dotted line:
HMRS(E), scale factor $p_t^{min}/2$. Experimental
data from UA1 (ref. [2]).
\vskip.1in
\item{Fig. 2}
Average number
of interactions, $\langle k(b=0, x)\rangle$, Eq.(11) in the text, as a function
of $x$ in nuclear collisions with sub-energy, in the
nucleon-nucleon c.m. system, $\sqrt s=1TeV$. Continuous lines: cut off
$p_t^{min}=5GeV$
and $p_t^{min}=6GeV$, HMRS(B) structure functions,
scale factor $p_t^{min}/2$. Dashed line: cut-off $p_t^{min}=5GeV$, HMRS(B)
structure functions,
scale factor $p_t^{min}$. Dotted line: HMRS(B) structure
functions, with shadowing corrections included,
cut-off $p_t^{min}=5GeV$, scale factor $p_t^{min}/2$.
Atomic mass number of the target $A=208$.
\vskip.1in
\item{Fig.3}
Dependence of $\langle k(b=0, x)\rangle$
on the nucleon-nucleon c.m. energy $\sqrt s$. Cut-off $p_t^{min}=5GeV$, HMRS(B)
structure functions,
scale factor $p_t^{min}/2$.
Solid line $x=1$, dashed line $x=0.5$, dotted line $x=0.25$. Atomic mass
$A=208$.
\vskip.1in
\item{Fig.4}
Dependence of $\langle k(b, x=1)\rangle$
on the impact parameter. Nucleon-nucleon c.m. energy $\sqrt s=1TeV$,
cut-off $p_t^{min}=5GeV$, HMRS(B) structure functions,
scale factor $p_t^{min}/2$,
atomic mass $A=208$.

\vfill
\eject\end
\bye